\documentstyle[12pt]{article}
 1

\advance \topmargin by -\headheight
\advance \topmargin by -\headsep

\evensidemargin \oddsidemargin
\begin{document}
\font\ninerm = cmr9

\def\footnoterule{\kern-3pt \hrule width \hsize \kern2.5pt}

\pagestyle{empty}
\begin{center}
{\large\bf Area-Preserving Structure of Massless}
{\large\bf Matter-Gravity Fields in 1+1 Dimensions}%
\footnote{\ninerm Talk given by G. Amelino-Camelia
at "Low-Dimensional Models in
Statistical Physics and Quantum Field Theory",
XXXIV Internationale Universit\"atswochen
f\"ur Kern- und Teilchenphysik,
Schladming, Austria,
March 4-11, 1995.
This work is supported
in part by funds provided by the U.S. Department of Energy (D.O.E.)
under cooperative agreement \#DE-FC02-94ER40818, as well as in part
by the National Science Foundation under contracts \#INT-910559 and
\#INT-910653, and by Istituto Nazionale di
Fisica Nucleare (INFN, Frascati, Italy). }

\vskip 1cm
G. Amelino-Camelia, D. Bak, and D. Seminara
\vskip 0.5cm
{\it Center for Theoretical Physics\\
Laboratory for  Nuclear Science and Department of Physics\\
Building 6, Massachusetts Institute of Technology\\
Cambridge, Massachusetts 02139, U.S.A.}

\end{center}

\vspace{3cm}
\begin{center}
{\bf ABSTRACT}
\end{center}

{\leftskip=0.6in \rightskip=0.6in

We derive anomalous Ward identities in two different
approaches to the quantization of massless
matter-gravity fields in 1+1 dimensions.
}

\vskip 2.5cm
$~$

\vfill
\hbox to \hsize{MIT-CTP-2438  \hfil May 1995}

\newpage
\baselineskip 12pt plus .5pt minus .5pt
\pagenumbering{arabic}
\pagestyle{plain}

\section{Introduction}
1+1 dimensional (``lineal") quantum gravity
is one of the areas of low-dimensional quantum field theory
which has attracted
more attention in recent years.
The role played by symmetries in these theories is obviously very
important, and here we shall be concerned
with the role of symmetries in the
quantization of the 1+1-dimensional matter-gravity field theory
with action\cite{pol81}
\begin{eqnarray}
{\cal I}(X,g) = {1 \over 2} \int d^2\xi ~ \sqrt{-g} ~
g^{\mu \nu} ~ \partial_\mu X_A \, \partial_\nu X^A ~,
\label{sxg}
\end{eqnarray}
where $g_{\mu \nu}$ is a metric tensor
with signature (1,-1), $A \! = \! 1,2,...,d$,
and $X$ is a d-component massless scalar field.

\noindent
In the quantization of this theory
one necessarily encounters anomalies that break part of
the symmetry of the classical theory, which, as seen from ${\cal I}(X,g)$,
has Weyl and diffeomorphism invariance\cite{pol81,pol87}.

In the conventional quantization
approach[1-4]
diffeomorphism invariance is preserved,
while  renouncing Weyl invariance.
Integrating out the
matter degrees of freedom using a
measure\cite{pol81,pol87} with the appropriate symmetries one obtains
an effective pure gravity theory
with action\footnote{Note that, for simplicity,
we set the cosmological constant to zero.}
\begin{eqnarray}
\Gamma^D(g) \! = {d \over 96 \pi} \!
\int \! d^2\xi_1 \, d^2\xi_2 \, \sqrt{-g(\xi_1)} \,
R(g(\xi_1)) ~ \Box^{-1}(\xi_1,\xi_2) ~ \sqrt{-g(\xi_2)} \, R(g(\xi_2))
\, , \label{sgp}
\end{eqnarray}
where $\Box^{-1}$ is the inverse of the Laplace-Beltrami operator.

Since $\Gamma^D(g)$ is diffeomorphism-invariant but
is not Weyl-invariant, the energy-momentum
tensor $T^D_{\mu \nu}  \! \equiv \! (2 / \sqrt{-g})
(\delta \Gamma^D(g) / \delta g^{\mu \nu})$
is covariantly conserved, but possesses non-vanishing trace
\begin{eqnarray}
\nabla_\mu (g^{\mu \nu} T^D_{\nu \alpha}) = 0  ~,
{}~~~~~g^{\mu \nu} T^D_{\mu \nu} = {d \over 24 \pi} R(g) ~.
\label{anop}
\end{eqnarray}

Recently, an alternative approach to the quantization of the classical
theory (\ref{sxg}) has been
considered[5-7],
in which the functional measure for the integration over the matter fields
is Weyl-invariant and invariant under area-preserving diffeomorphisms
({\it i.e.} diffeomorphisms of unit Jacobian),
but is not invariant under non-area-preserving diffeomorphisms.
This leads to the following
effective action\footnote{Note that, by appropriate choice
of measure, one can obtain more general
effective actions that are invariant
when $g_{\mu \nu}$ is transformed as
$\delta g_{\mu \nu} \! = \!
\xi^\alpha \partial_\alpha g_{\mu \nu}
+ g_{\alpha \nu} \partial_\mu \xi^\alpha
+ g_{\alpha \mu} \partial_\nu \xi^\alpha
+ a g_{\mu \nu} \partial_\alpha \xi^\alpha$,
where $a$ is a fixed real parameter.
It is seen that this combination of diffeomorphisms and
Weyl transformations
is equivalent to the
statement that $g_{\mu \nu}$ is a tensor density of
weight $a$.
Such a modification of the standard formula ($a \! = \!0$) leaves the
classical action invariant because
the combination $\sqrt{-g} g^{\mu \nu}$
is insensitive to the weight of $g^{\mu \nu}$.
The Weyl-invariant approach considered in the present paper
corresponds to the limit $a \! \rightarrow \! \infty$ with the
prescription that
$\partial_\mu \xi^\mu \! \rightarrow \! w/a$
for $a \! \rightarrow \! \infty$, where $w$
is an arbitrary function.}
\begin{eqnarray}
\Gamma^W(\gamma) \! = \! {d \over 96 \pi}
\int d^2\xi_1 \, d^2\xi_2 ~  \, R(\gamma(\xi_1)) \,
\Box^{-1}(\xi_1,\xi_2) \,  R(\gamma(\xi_2)) ~,
\label{sgj}
\end{eqnarray}
where $\gamma^{\mu \nu} \equiv \sqrt{- g} \, g^{\mu \nu}$.
The fact that $\Gamma^W(\gamma)$
is Weyl-invariant but is not invariant under general diffeomorphisms
leads to the anomaly relations
\begin{eqnarray}
\hat\nabla_\mu (\gamma^{\mu \nu} T^W_{\nu \alpha}) \!&=&\!
- {d \over 48 \pi} \partial_\alpha R(\gamma) ~,\label{anojgammad}\\
\gamma^{\mu \nu} T^W_{\mu \nu} \!&=&\! 0 ~,
\label{anojgammaw}
\end{eqnarray}
where $\hat\nabla$ is the
covariant derivative computed with the metric $\gamma_{\mu\nu}$, and
\begin{eqnarray}
T^W_{\mu \nu} ~ \equiv ~ {2 \over \sqrt{-g}}
{\delta \Gamma^W(\gamma) \over \delta g^{\mu \nu}}
{}~ = ~ 2 {\delta \Gamma^W(\gamma) \over \delta \gamma^{\mu \nu}} -
\gamma_{\mu \nu} \gamma^{\alpha \beta}
{\delta \Gamma^W(\gamma) \over \delta \gamma^{\alpha \beta}} ~,
\label{tj}
\end{eqnarray}
while the invariance of $\Gamma^W(\gamma)$ under area-preserving
diffeomorphisms is encoded in the relation
\begin{eqnarray}
\hat\nabla_\mu \hat\nabla_\nu (\gamma^{\beta \nu} \epsilon^{\mu \alpha}
T^W_{\alpha \beta}) = 0~,
\label{tjack}
\end{eqnarray}
which is consistent with (\ref{anojgammad}).

In the following we shall
derive the anomalous
Ward identities both for the
Weyl-invariant approach and
the conventional diffeomorphism invariant approach,
and observe that, although the difference in the symmetries
leads to several differences at intermediate steps of the
derivation, the final results are equivalent.

\section{Anomalous Ward Identities}
We start by considering the functional integrals
\begin{eqnarray}
Z^D[J] \!\!&=&\!\! \int {{\cal D} g \over \Omega_{diff}} ~
{\rm exp} (i \Gamma^D (g) + i \int \sqrt{-g} g^{\mu \nu} J^D_{\mu \nu}) ~,
\label{func}\\
Z^W[J] \!\!&=&\!\! \int {{\cal D} \gamma \over \Omega_{Sdiff}} ~
{\rm exp}(i \Gamma^W (\gamma)+ i \int \gamma^{\mu \nu} J^W_{\mu \nu}) ~,
\label{functional}
\end{eqnarray}
where $J^D_{\mu \nu}$ and $J^W_{\mu \nu}$ are sources,
$\Omega_{diff}$ is the volume of the diffeomorphism group,
and $\Omega_{Sdiff}$ is  the volume of the group of the area-preserving
diffeomorphisms. The volume of the Weyl
group does not appear in (\ref{functional}) because the functional
integral is already
written in terms of the Weyl-invariant field $\gamma$.

\noindent
$Z^D[J]$ and $Z^W[J]$ are the generating functionals for the
Green's functions of the diffeomorphism-invariant approach
and the Weyl-invariant approach
respectively.

In order to factorize out the gauge volume one can
fix the gauge
and introduce the corresponding action
for the ghost fields.
We choose
to work in the light-cone gauge, and,
after integrating out the ghost fields, $Z^{D}$ and $Z^{W}$ take
the following form
\begin{eqnarray}
\label{funct}
Z^D[J] \!\!&=&\!\! \int  {\cal D} g_{++} ~ {\rm exp}
\left (i \Gamma^D(g) + i \Gamma^D_{gh}(g) +
i \int g^{++} J^D_{++} \right ) ~, \label{functeff}\\
Z^W[J] \!\!&=&\!\! \int {\cal D} \gamma_{++} ~
{\rm exp} \left (i \Gamma^W({\gamma}) + i \Gamma^W_{gh}({\gamma})
+ i \int \gamma^{++} J^W_{++} \right )
{}~, \label{functional3}
\end{eqnarray}
where $\Gamma^{D} \! + \! \Gamma^{D}_{gh}$
and $\Gamma^{W} \! + \! \Gamma^{W}_{gh}$
are gauge-fixed actions for gravity.

\noindent
Our choice of gauge is motivated by the fact
that[5-7]
in the light-cone gauge $\Gamma^W \! + \! \Gamma^{W}_{gh}$
takes the same form of $\Gamma^D \! + \! \Gamma^{D}_{gh}$,
and we intend to exploit this correspondence
in the investigation of the anomalous Ward identities.
Still, in the analysis we shall need to take into account
the fact that the measure ${\cal D}\gamma$, which is Weyl-invariant
but not diffeomorphism-invariant, is different from
the diffeomorphism-invariant but not Weyl-invariant
measure ${\cal D}g$. Moreover, since
$\gamma$ and $g$ have different
transformation properties,
$\Gamma^W \! + \! \Gamma^{W}_{gh}$ and $\Gamma^D \! + \! \Gamma^{D}_{gh}$
satisfy different anomaly relations;
specifically\cite{abs1}
\begin{eqnarray}
\nabla_\mu (g^{\mu \nu} \Theta^D_{\nu \alpha}) = 0  ~,
{}~~~~~g^{\mu \nu} \Theta^D_{\mu \nu} = {d-28 \over 24 \pi} R(g) ~,
\label{anopgh}\\
\hat\nabla_\mu (\gamma^{\mu \nu} \Theta^W_{\nu \alpha}) =
- {d-28 \over 48 \pi} \partial_\alpha R(\gamma) ~,
{}~~~~~\gamma^{\mu \nu} \Theta^W_{\mu \nu} = 0 ~,
\label{anojgh}
\end{eqnarray}
where
\begin{eqnarray}
\Theta^{D,W}_{\mu \nu}  \equiv {2 \over \sqrt{-g}}
{\delta (\Gamma^{D,W} + \Gamma^{D,W}_{gh}) \over \delta g^{\mu \nu}} ~.
\label{tpgh}
\end{eqnarray}

We now consider
the following infinitesimal shifts in the functional variables
of integration
\begin{eqnarray}
\delta_f g_{++}=(2 \partial_+ -g_{++}\partial_-) \, \delta f +
\delta f \, \partial_- g_{++} ~,
\label{variation}\\
\delta_f \gamma_{++}=(2 \partial_+ -\gamma_{++}\partial_-) \, \delta f +
\delta f \, \partial_- \gamma_{++} ~,
\label{variationj}
\end{eqnarray}
and observe that
\begin{eqnarray}
\! \int \!
\frac{\delta [\Gamma^D(g) \! + \! \Gamma^D_{gh}(g)]}{\delta g_{++}}
\delta_f g_{++} \! = \!
\!\int \!
[ \nabla_\mu (g^{\mu \nu} \Theta^D_{\nu -})
\! - \! \frac{1}{2}\nabla_- (g^{\mu \nu} \Theta^D_{\mu \nu}) ]
\delta f
\! = \!\! \int \!
\frac{28 \! - \! d}{48 \pi}  \partial^3_- g_{++}  \delta f
\label{eqgac426}\\
\! \int \!
\frac{\delta [\Gamma^W(\gamma) \!
+ \! \Gamma^W_{gh}(\gamma)]}{\delta \gamma_{++}}
\delta_f \gamma_{++} \! = \!
\! = \!\! \int \!
[ \nabla_\mu (g^{\mu \nu} \Theta^W_{\nu -})
\! - \! \frac{1}{2}\nabla_- (g^{\mu \nu} \Theta^W_{\mu \nu}) ]
\delta f \! = \!\! \int \!
\frac{28 \! - \! d}{48 \pi} \partial^3_- \gamma_{++}  \delta f ,
\label{eqgac426bis}
\end{eqnarray}
where we used the anomaly relations (\ref{anopgh}) and (\ref{anojgh}).

Following a standard procedure\cite{bil95},
the relations (\ref{eqgac426}) and (\ref{eqgac426bis})
lead to the following
anomalous Ward identities
\begin{eqnarray}
\sum^n_i  \! \left < g_{++} \!  (\xi_1 ) \!
\dots \! \delta_{\!f} \! g_{++} \! (\xi_i ) \!
\dots  \! g_{++} \! (\xi_n)  \right > \!
+ \! \frac{d  \! \! - \! \!  28  \! \! +  \! \! \lambda^D}{i 48 \pi}
 \! \! \int \!  \! d\xi^2  \! \, \delta \! f \! ( \! \xi \! )
\left<  \partial_-^3 g_{++} \! (  \xi  )
g_{++} \! (\xi_1 ) \! \dots  \! g_{++} \!
(\xi_n)  \right > \!  \! = \!  \! 0.
\label{Ward}\\
\sum^n_i  \! \left < \gamma_{++} \!  (\xi_1 ) \!
\dots \! \delta_{\!f} \! \gamma_{++} \! (\xi_i ) \!
\dots  \! \gamma_{++} \! (\xi_n)  \right > \!
+ \! \frac{d  \! \! - \! \!  28  \! \! +  \! \! \lambda^W}{i 48 \pi}
 \! \! \int \!  \! d\xi^2  \! \, \delta \! f \! ( \! \xi \! )
\left<  \partial_-^3 \gamma_{++} \! (  \xi  )
\gamma_{++} \! (\xi_1 ) \! \dots  \! \gamma_{++} \!
(\xi_n)  \right > \!  \! = \!  \! 0.
\label{Wardj}
\end{eqnarray}
Here $\lambda^D$ is the additional contribution to the anomaly
which is due to the fact that $\delta_f g_{++}$
is a composition of a diffeomorphism
and a Weyl transformation on $g_{++}$,
and therefore the diffeomorphism-invariant
but not Weyl-invariant measure ${\cal D}g_{++}$ is not invariant
under $g_{++} \rightarrow g_{++} + \delta_f g_{++}$.
Analogously, the presence of $\lambda^W$ is due to the fact that
$\delta_f \gamma_{++}$ is an infinitesimal
(not area-preserving) diffeomorphism
transformation on $\gamma_{++}$,
and therefore the measure ${\cal D}\gamma_{++}$ is not invariant
under $\gamma_{++} \rightarrow \gamma_{++} + \delta_f \gamma_{++}$.
The values of $\lambda^{D}$ and $\lambda^{W}$ can
be fixed by requiring that
the theory be independent of the choice of gauge.
In Ref.\cite{pol87}
the class of gauges
$g_{--} \! = \!g_{--}^B$, $g_{+-} \! = \! 1$ is considered, and
it is found that the independence of the partition function on
the choice of $g_{--}^B$ requires that
\begin{equation}
\label{lambda}
d-28+\lambda^D = {d-13- \sqrt{(d -1)(d - 25)} \over 2}~.
\end{equation}
Following the corresponding procedure for the Weyl invariant approach
one finds that also $\lambda^W$ must satisfy Eq.(\ref{lambda}),
{\it i.e.} $\lambda^W \! = \! \lambda^D$.
This observation together with the results (\ref{Ward}) and (\ref{Wardj})
indicates that the anomalous Ward identities satisfied by $\gamma_{++}$
in the Weyl-invariant approach are identical to the ones
satisfied by $g_{++}$
in the diffeomorphism-invariant approach.
Since these Ward identities completely
determine\cite{pol87}
the Green's functions, also the Green's functions
are identical.

\section{Conclusion}
The investigation of the anomalous Ward identities
indicates that the two approaches are equivalent,
and this is consistent with the results\cite{abs1} of the (classical)
Dirac Hamiltonian analysis.
It appears that the physics described by the model is independent
of the local term\cite{jac95} that one needs to add to the action
in order to convert the Weyl anomaly into a diffeomorphism anomaly.
This does not always happen in anomalous quantum field theories,
for example in the chiral Schwinger model\cite{jac85}
the mass emergent
at the quantum level as a result of the anomaly
does depend on the coefficient of one such local term.

It is also interesting to notice that
in deriving the equivalence of the two approaches at the level of
the anomalous Ward identities
a key role is played by the $-$ component of the combination
$\nabla_\mu (g^{\mu \nu} \Theta_{\nu \alpha})
-\nabla_\alpha (g^{\mu \nu} \Theta_{\mu \nu})/2$
[see Eqs.(\ref{eqgac426}) and (\ref{eqgac426bis})],
which (in the chosen gauges) takes the same form in both approaches.
Clearly this combination of the anomaly relations
encodes some essential feature of the model,
but its physical interpretation is not yet clear to us.

Finally, we want to point out that
1+1-dimensional quantum gravities of the type here consedered
and their supersymmetric extensions are related\cite{pol87,dis88,stat}
to some low-dimensional models in statistical physics,
such as the Ising model, random surfaces, percolation, tree-like polymers,
and self-avoiding polymers.
In several occasions
results first obtained in the study of the quantum field theories have
been useful also in the context of the statistical models and
{\it vice versa}.
Since only recently there has been increased interest in the Weyl invariant
approach, the possibility of
application of this new viewpoint to the study of statistical
models has not yet been investigated.

\vglue 0.6cm
\leftline{\Large {\bf Acknowledgements}}
\vglue 0.3cm
We are very grateful to R. Jackiw for several stimulating conversations
and useful comments.

\baselineskip 12pt plus .5pt minus .5pt

\end{document}